\documentstyle[12pt]{article}
\begin{document}
\title{WEAK HYPERNUCLEAR DECAY}
\author{Barry R. Holstein\\
{\em Department of Physics and Astronomy}\\
{\em University of Massachusetts}\\
{\em Amherst, MA  01003}\\
and\\
{\em Institute for Nuclear Theory}\\
{\em Department of Physics, NK-12}\\
{\em University of Washington}\\
{\em Seattle, WA 98195}}
\maketitle
\begin{abstract}
{Because of Pauli suppression effects the $N\pi$ decay mode of the free
$\Lambda$
is not of importance in hypernuclei with $A\geq 10$.  Rather the decay of such
hypernuclei proceeds via the nucleon-stimulated mode
$\Lambda N\rightarrow NN$, analysis of which presents a
considerable theoretical challenge and about which there exists only a
limited amount of experimental information.  Herein we confront existing
data with various
theoretical analyses which have been developed.}
\end{abstract}
\section{-- Introduction to Hypernuclear Decay}
The properties of the lambda hyperon are familiar to all of us.  Having a mass
of 1116 MeV, zero isospin and unit negative strangeness, it decays nearly 100\%
of the time
via the nonleptonic mode $\Lambda\rightarrow N\pi$ and details can be
found in the particle data tables\cite{1}
\begin{equation}
\Gamma_\Lambda={1\over 263 \mbox{ps}}\qquad \mbox{B.R.}\,\Lambda\rightarrow
\left\{\begin{array}{cc}
p\pi^- & 64.1\% \\
n\pi^0 & 35.7\%
\end{array}\right.
\end{equation}
The decay can be completely described in terms of an effective Lagrangian with
two phenomenological parameters
\begin{equation}
{\cal H}_w=g_w\bar{N}(1+\kappa\gamma_5)\vec{\tau}\cdot
\vec{\phi}_\pi\Lambda
\end{equation}
where $g_w=2.35\times 10^{-7},\kappa=-6.7$ and $\Lambda$ is defined to occupy
the
lower entry of a two component column spinor.  The kinematics are such that
for decay at rest the final state nucleon emerges with energy about 5 MeV,
which means that the corresponding momentum is $p_N=\sqrt{2M_NE_N}\approx
100$ MeV.

Now, however, consider what happens if the Lambda is bound in a hypernucleus.
\cite{2}
In this case, even neglecting binding energy effects, the 100 MeV momentum of
the outgoing nucleon is generally much less than the Fermi momentum of the
nucleus so that the decay will be Pauli blocked.  A very simple estimate of
this effect can be generated within a simple Fermi gas model, wherein,
neglecting any effects of binding energy or of wavefunction distortion, one
finds
\begin{equation}
{1\over \Gamma_\Lambda}\Gamma_{\pi}=1-{1\over 2}\sum_{nj\ell}N_{nj\ell} \langle
nj\ell|j_\ell(k_\pi r)|1S_{1\over 2}\rangle
\end{equation}
with $N_{nj\ell}$ being the occupation number for the indicated state.  The
result of this calculation reveals that the importance of such pionic decays
rapidly falls as a function of nuclear mass---
\begin{equation}
{1\over \Gamma_\Lambda}\Gamma_{\pi}\sim\left\{
\begin{array}{ccc}
A=10 & A=25 & \ldots \\
1/20 & 1/120 & \ldots
\end{array}\right.
\end{equation}
However, while the existence of the nuclear medium suppresses the
$N\pi$ mode, it also opens
up a completely new possibility, that of the nucleon-stimulated
decay---$\Lambda N\rightarrow NN$.
Assuming that the energy is shared equally between the
outgoing pair of nucleons one has then $E_N\simeq {1\over 2}(m_\Lambda-m_N)
\approx 90 MeV$.  The corresponding momentum is
$p_N\sim 400 MeV$ and is well above the Fermi momentum, so that Pauli
suppression is not relevant.  According to the above arguments the
importance of this nonmesonic (NM) mode compared to its mesonic
counterpart should rapidly increase with A, and this expectation is fully borne
out experimentally, as shown in Figure 1.
\begin{figure}
\vspace{3.5in}
\caption{Calculated ratio of pionic hypernuclear decay to free lambda decay
rate.}
\end{figure}
A theory of hypernuclear weak decay then has basically nothing to do with the
pionic mode favored by a free $\Lambda$ and must deal with the much more
complex $\Lambda N\rightarrow NN$ process.\cite{3}  The observables which can
be
measured experimentally and should be predicted by theoretical analysis include
\begin{itemize}
\item[i)] the overall decay rate $\Gamma_{NM}$;
\item[ii)] the ratio of proton-stimulated ($\Lambda p\rightarrow np$) to
neutron-stimulated ($\Lambda n\rightarrow nn$)
decay---$\Gamma^p_{NM}/\Gamma^n_{NM}
\equiv \Gamma_{NM}(p/n)$;
\item[iii)] the ratio of parity-violating to parity-conserving
decay---
$$\Gamma^{PV}_{NM}/\Gamma^{PC}_{NM}\equiv\Gamma_{NM}(PV/PC)$$
---which is measured, {\it e.g.},
via the proton asymmetry in polarized hypernuclear decay
\item[iv)] final state n,p decay spectra;
\item[v)] {\it etc.}
\end{itemize}

The present experimental situation is somewhat limited. Most of the early
experiments in the field employed bubble chamber or emulsion techniques.
It was therefore relatively straightforward to determine the ratio of the
decay rates of the two modes, but much more difficult to measure the
absolute rates.  This changed when an early Berkeley
measurement on ${}^{16}_\Lambda \mbox{O}$ yielded the value\cite{4}
\begin{equation}
{\Gamma({}^{16}_\Lambda \mbox{O})\over \Gamma_\Lambda}=3\pm 1
\end{equation}
However, this was still a very low statistics experiment with sizable
background contamination.  Recently a CMU-BNL-UNM-Houston-Texas-Vassar
collaboration undertook a
series of  direct timing---fast counting---hypernuclear lifetime measurements
yielding the results summarized in table 1.\cite{5}
In addition, there exist a number of older emulsion measurements in light
($A\leq 5$) hypernuclei, details of which can be found in a recent review
article.\cite{6}  However, the only experimental numbers for heavy
systems are obtained from delayed fission measurements on hypernuclei
produced in $\bar{p}$-nucleus collisons and are of limited statistical
precision\cite{7}
\begin{equation}
\Gamma({}^{238}_{\Lambda}\mbox{U})=(1.0^{+1.0}_{-0.5})\times 10^{-10}
\mbox{sec.}\qquad
\Gamma({}^{209}_{\Lambda}\mbox{Bi})=(2.5^{+2.5}_{-1.0})\times
10^{-10}\mbox{sec}.
\end{equation}
\begin{table}
\begin{center}
\begin{tabular}{|c|c|c|c|}
\hline
  &${}^5_\Lambda \mbox{He}$ &$ {}^{11}_\Lambda \mbox{B}$ & ${}^{12}_\Lambda
\mbox{C} $\\
\hline
\hline
${1\over \Gamma_\Lambda}\Gamma_{NM}$ & $0.41\pm 0.14$& $ 1.25\pm 0.16^*$ &
$1.14\pm 0.20$ \\ \hline
$\Gamma_{NM}(p/n)$ &$1.07\pm 0.58 $  &$0.96^{+0.8}_{-0.4}$
&$0.75^{+1.5}_{-0.35}$ \\ \hline
\end{tabular}
\end{center}
\caption{Experimental BNL data for nonmesonic hyperon decay.
$^*$Note that we have scaled the experimental number to account to exclude the
pionic decay
component.}
\end{table}

The problem of dealing
with a weak two-body interaction within the nucleus has been faced previously
in the
context of nuclear parity violation, and one can build on what has been learned
therein.\cite{8}  Specifically, the weak interaction at the quark level is
shortranged, involving W,Z-exchange.  However, because of the hard core
repulsion the
effective NN effects are modelled in terms of long-range one-meson
exchange interaction, just as in the case of the conventional strong
nucleon-nucleon interaction,\cite{9} but now
with one vertex being weak and parity-violating while
the second is strong and parity-conserving.  The exchanged mesons are
the lightest ones---$\pi^\pm ,\rho ,\omega$---associated with the longest
range. (Exchange of
neutral spinless mesons is forbidden by Barton's theorem.\cite{10})
\begin{figure}
\vspace{2.5in}
\caption{Meson exchange picture of nuclear parity violation.}
\end{figure}

A similar picture of hypernuclear decay can then be constructed, but with
important differences.  While the basic meson-exchange diagrams appear as
before, the weak vertices must now include both parity-conserving {\it and}
parity-violating components, and the list of exchanged mesons must be expanded
to include
both neutral spinless objects ($\pi^0,\eta^0$) as well as strange mesons
($K,K^*$), as first pointed out by Adams.\cite{11}  Thus the problem is
considerably more
challenging than the corresponding and already difficult issue of nuclear
parity violation.

One of the significant problems in such a calculation involves the evaluation
of the various
weak amplitudes.  Indeed, the only weak couplings which are completely
model-independent are those involving pion emission, which are given in Eqn. 2.
In view of this, a number of calculations have included {\it only} this
longest range component.
Even in this simplified case, however,
there is considerable model-dependence, as the results are
strongly sensitive to the short-ranged correlation function assumed for the
nucleon-nucleon interaction, as will be seen.
Below we shall review previous theoretical work in this area and detail our
own program, which involves a systematic quark model- (symmetry-) based
evaluation of weak mesonic couplings to be used in hypernuclear decay
calculations.

\section{-- Hypernuclear Decay in Nuclear Matter}

As discussed above one of the significant problems in the calculation of
hypernuclear decay involves the evaluation of the various
weak NNM vertices.  Indeed, the only weak couplings which are completely
model-independent are those involving pion emission, which are given in Eqn. 2.
In view of this, a number of calculations have included {\it only} this longest
range
component.
Even here there is considerable model-dependence, however, as the results are
strongly sensitive to the short-ranged correlation function assumed for the
nucleon-nucleon interaction.  As a warm-up to a realistic calculation then we
can begin with a pion-exchange-only calculation in
``nuclear matter" ({\it i.e} a simple Fermi gas model with $N_n=N_p$ and
$P_f\sim 270 $ MeV) with and without nucleon-nucleon correlation effects.  Here
the
$\Lambda -N$ relative momentum is very soft so that only
${}^1S_0$ and ${}^3S_1$ initial states are assumed to be involved.  Then
\begin{eqnarray}
\Gamma_{NM}&=&{1\over (2\pi)^5}\int d^3k_1\int
d^3k_2\int_0^{k_F}d^3k_i\delta^4(p_i-p_f)\nonumber\\
&\times &{1\over 2}\sum_{\rm spin}|\langle f|{\cal
H}_w|i\rangle|^2=\sum_{\alpha\beta}\Gamma_{NM}(\beta\leftarrow\alpha)
\end{eqnarray}

\begin{table}
\begin{center}
\begin{tabular}{|l|l|}\hline
Transition & Operator \\
\hline
\hline
${}^1S_0\rightarrow{}^1S_0(I=1)$ & ${1\over 4}a(q^2)
(1-\vec{\sigma}_\Lambda\cdot\vec{\sigma}_N)$\\ \hline
${}^1S_0\rightarrow{}^3P_0(I=1)$ &$ {1\over 8}b(q^2)(\vec{\sigma}_\Lambda
-\vec{\sigma}_N)\cdot\hat{q}(1-\vec{\sigma}_\Lambda\cdot
\vec{\sigma}_N)$\\ \hline
${}^3S_1\rightarrow{}^3S_1(I=0)$ &$ {1\over 4}c(q^2)(3+\vec{\sigma}\cdot
\vec{\sigma}_N)$\\
\hline
${}^3S_1\rightarrow{}^3D_1(I=0)$ &$ {3\over 2\sqrt{2}}d(q^2)
(\vec{\sigma}_\Lambda\cdot\hat{q}\vec{\sigma}_N\cdot
\hat{q}-{1\over 3}
\vec{\sigma}_\Lambda\cdot\vec{\sigma}_N)$ \\ \hline
${}^3S_1\rightarrow{}^1P_1(I=0)$ &$ {\sqrt{3}\over 8}e(q^2)(\vec{\sigma}
_\Lambda
-\vec{\sigma}_N)\cdot\hat{q}(3+\vec{\sigma}_\Lambda\cdot
\vec{\sigma}_N)$ \\ \hline
${}^3S_1\rightarrow{}^3P_1(I=1)$ &$ {\sqrt{6}\over 4}f(q^2)(\vec{\sigma}
_\Lambda
+\vec{\sigma}_N)\cdot\hat{q}$ \\ \hline
\end{tabular}
\end{center}
\caption{Transition operators of allowed $\Lambda N\rightarrow NN$ transitions
from relative S-states.  Here $\vec{q}$ specifies the relative momentum of the
outgoing nucleons while $\vec{\sigma}_\Lambda,\vec{\sigma}_N$
operate on the $\Lambda N,NN$ vertices respectively.}
\end{table}
We can break this down further by identifying effective transition operators
for the various partial wave channels which contribute to the decay process
---{\it cf.} Table 2---in terms of which we find for the total nonmesonic
hypernuclear decay rate
\begin{equation}
\Gamma_{NM}={3\over 8\pi\mu^3_{\Lambda N}}\int_0^{\mu_{\Lambda N}k_F}
p^2dpqm_N\left(|a|^2+|b|^2+|c|^2+|d|^2+|e|^2+3|f|^2\right)
\end{equation}
where $\mu_{\Lambda N}={m_\Lambda\over m_\Lambda+m_N}$ arises from the switch
from the nuclear rest frame to the $\Lambda N$ center of mass frame and p,q
are related by
\begin{equation}
q^2=m_N(m_\Lambda-m_N)+{p^2\over 2\mu_{\Lambda N}}
\end{equation}
The results obtained by various groups are displayed in Table 3.
\begin{table}
\begin{center}
\begin{tabular}{|c|c|c|c|c|}\hline
   & Adams\cite{11} & McK-Gib\cite{12} & Oset-Sal\cite{13} & UMass\cite{14} \\
\hline
\hline
${1\over \Gamma_\Lambda}\Gamma_{NM}$(no corr.)& 0.51 & 4.13 &4.3 & 3.84 \\
\hline
${1\over \Gamma_\Lambda}\Gamma_{NM}$(corr.) & 0.06 & 2.31 & 2.1 & 1.82 \\
\hline
\end{tabular}
\end{center}
\caption{Non-mesonic hypernuclear decay rates calculated by various groups
using pion-exchange only in ``nuclear matter."}
\end{table}

Obviously there is basic agreement except for the pioneering calculation of
Adams.\cite{11}   The problems with his calculation were two---Adams
used an incorrect value of the weak
coupling constant $g_w$ as well as too-strong a tensor correlation,
both of which tended to reduce the calculated rate.  When these are corrected
the corresponding numbers are found to be 3.5 (no correlations) and 1.7 (with
correlations) and are in basic agreement with other predictions.
{}From this initial calculation then we learn that the basic nonmesonic decay
rate is indeed anticipated to be of the same order as that for the free
$\Lambda$ {\it and} the important role played by correlations.

A second quantity of interest which emerges
from such a calculation is the p/n stimulated decay ratio, given by
\begin{equation}
\Gamma_{NM}(p/n)={\int_0^{\mu_{\Lambda N}k_F}p^2dpq(|a|^2+|b|^2+3|c|^2+3|d|^2
+3|e|^2+3|f|^2)\over \int_0^{\mu_{\Lambda}k_F}p^2dpq(2|a|^2+2|b|^2+6|f|^2)}
\end{equation}
and which has been calculated by two of the groups, yielding the results shown
in Table 4.
\begin{table}
\begin{center}
\begin{tabular}{|c|c|c|c|c|}\hline
    & Adams\cite{11} & McK-Gib\cite{12} & Oset-Sal\cite{13} & UMass\cite{14} \\
\hline
\hline
$\Gamma_{NM}(p/n)$ (no corr.) & 19.4 & - & -& 11.2 \\ \hline
$\Gamma_{NM}(p/n)$ (corr.) & 2.8  & - & - & 16.6 \\ \hline
\end{tabular}
\end{center}
\caption{Proton to neutron stimulated decay ratios for pion-only exchange in
``nuclear matter."}
\end{table}
An interesting feature here is that the numbers come out so large---proton
stimulated decay is predicted to predominate over its
neutron stimulated counterpart by nearly an order of magnitude.
The reason for this is easy to see.  In a pion-exchange-only scenario the
effective weak interaction is of the form
\begin{equation}
{\cal H}_w\sim g\bar{N}\vec{\tau} N\cdot \bar{N}\vec{\tau} \Lambda
\end{equation}
Then $\Lambda n\rightarrow nn\sim g$ but $\Lambda p\rightarrow np\sim (-1-
(\sqrt{2})^2)g=-3g$ since both charged and neutral pion exchange are involved.
In this naive picture then we have $\Gamma_{NM}(p/n)\sim 9$, in rough
agreement with the numbers given in Table 4.

Armed now with theoretical expectations, we can ask what does experiment say?
The only reasonably precise results obtained for nuclei with $A>4$
are those
measured at BNL on ${}^5_\Lambda \mbox{He},\quad
{}^{12}_\Lambda \mbox{C}$ and ${}^{11}_\Lambda \mbox{B}$, which are
summarized in Table 1.  We observe that the measured nonmesonic decay
rate is about a factor of two lower than that predicted in Table 3
while the p/n stimulation ratio differs
by at least an order of magnitude from that given in Table 4.  The problem
may be, of course, associated with the difference between the nuclear
matter within which the calculations were performed and the finite
nuclear systems which were examined experimentally.  Or it could be due to
the omission of the many shorter range exchanged mesons in the theoretical
estimate. (Or {\it both!})

Before undertaking the difficult problem of
finite nuclear calculations, it is useful to first examine the inclusion of
additional
exchanged mesons in our calculations.  As mentioned above, a primary difficulty
in this approach
is that none of the required weak couplings can be measured experimentally.
Thus the use of some sort of model is required, and the significance of any
theoretical predictions will be no better than the validity of the model.
One early attempt by McKellar and Gibson, for
example, included only the rho and evaluated the rho couplings using both SU(6)
symmetry methods as well as the well
known but flawed factorization approach.\cite{12}  Well aware of the
limitations of
this method, they allowed an arbitrary phase between the rho and pi amplitudes
and they renormalized the factorization calculation by a factor of
$1/\sin\theta_c\cos\theta_c$ in order to account for the $\Delta I={1\over 2}$
enhancement.  Obviously this is only a rough estimate then and this is only for
the
rho meson exchange contribution!  A similar approach was attempted by
Nardulli, who calculated the parity conserving rho amplitude in a simple
pole model and the parity violating piece in a simple quark picture.\cite{15}
Results of these calculations are shown in Table 5
\begin{table}
\begin{center}
\begin{tabular}{|c|c|c|c|}\hline
    & McK-Gib\cite{12} $\pi+\rho$ & McK-Gib\cite{12} $\pi-\rho$ &
Nard.\cite{15} \\
\hline
\hline
${1\over \Gamma_\Lambda}\Gamma_{NM}$ & 3.52 & 0.72 & 0.7 \\ \hline
\end{tabular}
\end{center}
\caption{Nonmesonic decay rates in nuclear matter in pi plus rho exchange
models}
\end{table}

To my knowledge, the only comprehensive calculation which has been undertaken
to date is that of our group at UMass.  In the case of the parity violating
interaction we employed a variant of the (broken) $SU(6)_w$
symmetry calculations
which were employed successfully by Desplanques, Donoghue and Holstein to
calculate the various weak NNM couplings in the case of nuclear parity
violation.\cite{8}  In this approach there exist three (in principle unknown)
reduced
matrix elements which, when multiplied by the relevant generalized
``Clebsch-Gordan"
coefficients, relate all such parity-violating amplitudes.  Two of these are
determined empirically in terms of experimental hyperon decay data, while the
third is given by a factorization calculation.  While the success of this
approach in the case of nuclear parity violation is not without
question,\cite{16}
this procedure provides a plausible and unambiguous approach to the
problem.

More difficult is the determination of the parity-conserving weak couplings.
In this case we employ a pole model using the diagrams shown in Figure 3.
\begin{figure}
\vspace{3in}
\caption{Pole diagrams used in evaluation of weak parity-conserving
$\Lambda N\rightarrow NN$ couplings.}
\end{figure}
What is needed here are the weak parity-conserving amplitudes for $\Lambda -N$
and $\Sigma -N$ transitions, which we determine via the current algebra
(chiral symmetry) relations
\begin{eqnarray}
\lim_{q_\pi\rightarrow 0}\langle\pi^0n|{\cal H}_w^{(-)}|\Lambda\rangle
&=&{-i\over F_\pi}\langle n|[F^5_{\pi^0},{\cal H}_w^{(-)}]|\Lambda\rangle =
{i\over 2F_\pi}\langle n|{\cal H}_w^{(=)}|\Lambda\rangle\nonumber\\
\lim_{q_\pi\rightarrow 0}\langle\pi^0p|{\cal H}_w^{(-)}|\Sigma^+\rangle
&=&{-i\over F_\pi}\langle p|[F^5_{\pi^0},{\cal H}_w^{(-)}|\Sigma^+\rangle =
{i\over 2F_\pi}\langle p|{\cal H_w^{(-)}}|\Sigma^+\rangle
\end{eqnarray}
and the weak $K-\pi$ coupling which is similarly given in terms of the
experimental $K\pi\pi$ decay amplitude
\begin{equation}
A_{K\pi}=-iF_\pi{k\cdot q\over m_K^2-m_\pi^2}\langle \pi^0\pi^0|{\cal
H}_w|K^0\rangle_{\rm physical}
\end{equation}
Again this procedure has well documented flaws.\cite{17}  However, in the
present context it is reasonable successful and
for a first generation calculation, we consider it to provide a
reasonable estimate for the parity conserving weak couplings.

Combining with the various strong meson couplings we can now substitute into
the diagrams shown in Figure 3 to generate the many effective parity conserving
two-body operators which can be used to evaluate the nonmesonic decay
amplitudes.
Details of these procedures are given in ref. 10.   Using the resultant
two-body operators one can then generate the various
predictions for nonmesonic decay in nuclear matter.  Results are
summarized in table 6, where we specifically
identify the contributions from various channels.
\begin{table}
\begin{center}
\begin{tabular}{|c|c|c|c|c|}\hline
   & $\pi$  & $\pi$ &     & $\pi,\rho,\eta$ \\
   & (no corr.) & (corr.) & $\pi+\rho$ & $\omega,K,K^*$ \\
\hline
\hline
${}^1S_0\rightarrow {}^1S_0$ & .010 & .000 & .001 & .001 \\ \hline
${}^1S_0\rightarrow {}^3P_0$ & .156 & .037 & .052 & .018 \\ \hline
${}^3S_1\rightarrow {}^3P_1$ & .312 & .117 & .113 & .456 \\ \hline
${}^3S_1\rightarrow {}^1P_1$ & .468 & .128 & .100 & .110 \\ \hline
${}^3S_1\rightarrow {}^3S_1$ & .010 & .789 & .589 & .202 \\ \hline
${}^3S_1\rightarrow {}^3D_1$ & 2.93 & .751 & .693 & .444 \\ \hline
Total                & 3.89 & 1.82 & 1.55 & 1.23 \\ \hline
\end{tabular}
\end{center}
\caption{Decay rates for various combinations of meson exchange in nuclear
matter.}
\end{table}

The results are very intriguing.  The overall decay rate is reduced somewhat
from its pion-exchange-only value,
in agreement with the experimental results.  More striking is the modification
of the p/n ratio
and in the ratio of parity violating to parity conserving decay, defined as
\begin{equation}
\Gamma_{NM}(PV/PC)={\int_0^{\mu_\Lambda N}p^2dpq(|b|^2+|e|^2+|f|^2)\over
\int_0^{\mu_{\Lambda N}}p^2dpq(|a|^2+|c|^2+3|d|^2)}
\end{equation}
values of which are shown in Table 7.
\begin{table}
\begin{center}
\begin{tabular}{|c|c|c|}\hline
    &$ \Gamma_{NM}(PV/PC)$ & $\Gamma_{NM}(p/n)$\\
\hline
\hline
$\pi$ (no corr.) &  0.14 & 11.2 \\ \hline
$\pi$ (with corr.) & 0.18 & 16.6 \\ \hline
$\pi+\rho $  & 0.21 & 13.1 \\ \hline
$\pi,\rho,\omega,\eta,K,K^*$ & 0.90 & 2.9 \\ \hline
\end{tabular}
\end{center}
\caption{The parity violating to parity conserving and p to n ratios for
hypernuclear decay in ``nuclear matter."}
\end{table}
We observe that inclusion of additonal exchanges plays a major role in reducing
the p/n ratio from its pion-only-exchange value.  The resulting value of
2.9 is still somewhat larger than the experimental values shown in Table 1
but clearly indicate the presence of non-pion exhange components.

The reason that kaon exchange in particular can play such a major role
can be seen from a simple argument due to Gibson\cite{18} who pointed out
that since the final NN system can have either I=0 or I=1, the effective
kaon exchange interaction can be written as
\begin{eqnarray}
{\cal L}_{eff}=A_0(\bar{p}p+\bar{n}n)\bar{n}\Lambda+A_1(2\bar{n}p\bar{p}\Lambda
-(\bar{p}p-\bar{n}n)\bar{n}\Lambda)\nonumber\\
\sim(A_0-3A_1)\bar{p}p\bar{n}\Lambda+(A_0+A_1)\bar{n}n\bar{n}\Lambda
\end{eqnarray}
where the second line is obtained via a Fierz transformation.  Since for
parity violating kaon exchange we have $A_0\sim 6A_1$ we find\cite{18}
\begin{equation}
\Gamma_{NM}(p/n)=\left({A_0-3A_1\over A_0+A_1}\right)^2\sim 1/5
\end{equation}
which clearly indicates the importance of inclusion of non-pion-exchange
components in predicting the p/n ratio.

A second strong indication of the presence of non-pion-exchange can
be seen from Table 7 in that the rate of parity violating to parity
conserving transitions is substantially enhanced by the inclusion of
kaon and vector meson exchange as compared to the simple pion-exchange-only
calculation.  We can further quantify this effect by calculating explicitly
the angular distribution of the emitted proton in the $\Lambda p\rightarrow np$
transition (there can be no asymmetry for the corresponding
$\Lambda n\rightarrow nn$ case due to the identity of the final state
neutrons),
yielding
\begin{equation}
W_p(\theta)\sim 1+\alpha P_\Lambda\cos\theta
\end{equation}
where
\begin{equation}
\alpha={\int_0^{\mu_{\Lambda N}k_F}p^2dpq{\sqrt{3}\over 2}{\rm Re}f^*
(\sqrt{2}c+d)\over \int_0^{\mu_{\Lambda N}k_F}p^2dpq{1\over 4}(|a|^2+|b|^2
+3|c|^2+3|d|^2+3|e|^2+3|f|^2}
\end{equation}
is the asymmetry parameter.  Results of a numerical evaluation are shown in
Table 8
\begin{table}
\begin{center}
\begin{tabular}{|c|c|c|c|}\hline
\quad & $\pi$-no corr. & $\pi$-corr. & all exch. \\
\hline
\hline
$\alpha$ & -0.078 & -0.192 & -0.443 \\ \hline
\end{tabular}
\end{center}
\caption{Proton asymmetry coefficient in various scenarios.}
\end{table}
so that again inclusion of non-pion-exchange components has a significant
effect, increasing the expected $\Lambda p\rightarrow np$ asymmetry
by more than a factor of two.  This prediction of a substantial asymmetry
is consistent with preliminary results obtained for p-shell nuclei at
KEK.\cite{19}

\section{-- Hypernuclear Decay in Finite Nuclei}

Although the nuclear matter calculations are of great interest in identifying
basic properties of the decay process, true confrontation with experiment
requires calculations involving the finite nuclei on which the
measurements are conducted.  Of course, such calculations are
considerably more challenging than their nuclear matter counterparts and
require $\Lambda$ shell model considerations as well as non-S-wave capture.
Nevertheless a number of groups have taken up the challenge.  For the case of
the nonmesonic decay of ${}^{12}_\Lambda$C the results are summarized in
Table 9.
\begin{table}
\begin{center}
\begin{tabular}{|c|c|c|c|}\hline
  & Oset-Sal\cite{13}& TRIUMF\cite{20} & UMass\cite{14} \\
\hline
\hline
${1\over \Gamma_\Lambda}\Gamma_{NM}\,\,\, \pi$ (no corr.)&  & 1.6 & 3.4 \\
\hline
$\qquad\pi$ (corr.) & 1.5 & 2.0 & 0.5\\ \hline
$\pi+K$\cite{20};\,\,\,$\pi,\eta,\rho\omega,K,K^*$\cite{14}& & 1.2 & 0.2 \\
\hline
$\Gamma_{NM}(p/n) \pi$ (no corr.)& & 5.0 & 4.6 \\ \hline
\qquad $\pi$ (corr.) & & 5.0 & 5.0 \\ \hline
$\pi+K$\cite{20};\,\,\,$\pi,\eta,\rho,\omega,K,K^*$\cite{14}& & 4.0 & 1.2 \\
\hline
$\Gamma_{NM}(PV/PC) \pi$ (no corr.) & & 0.4 & 0.1 \\ \hline
\qquad $\pi$ (corr.) & & 0.5 & 0.1 \\ \hline
$\pi+K$\cite{20};\,\,\,$\pi,\eta,\rho,\omega,K,K^*$\cite{14}& & 0.3 & 0.8 \\
\hline
\end{tabular}
\end{center}
\caption{Calculated properties of nonmesonic hypernuclear decay of
${}^{12}_\Lambda$C.}
\end{table}
In comparing with the experimental results given in Table 1, we see that the
UMass calculation is certainly satisfactory, but the discrepancy between the
UMass and TRIUMF work is disturbing and needs to be rectified before either
is to be believed.

A second nucleus on which there has been a good deal of work, both
experimentally and theoretically, is ${}^5_\Lambda$He, which is summarized
in Table 10.
\begin{table}
\begin{center}
\begin{tabular}{|c|c|c|c|c|}\hline
    & Oset-Sal\cite{13}&TRIUMF\cite{20}&TTB\cite{21}&UMass\cite{14}\\
\hline
\hline
${1\over \Gamma_\Lambda}\Gamma_{NM} \pi$(no corr.)&  & 1.0 & 0.5 & 1.6 \\
\hline
$\pi$(corr.) & 1.15 & 0.25 & 0.144 & 0.9 \\ \hline
$\pi+K$\cite{20};\,\,\,$\pi,\eta,\rho,\omega,K,K^*$\cite{14}& & 0.22 & & 0.5\\
\hline
$\Gamma_{NM}(p/n) \pi$ (no corr.)&  & 5.0 & & 15 \\ \hline
$\pi$ (corr.)& & 4.8& & 19 \\ \hline
$\pi+K$\cite{20};\,\,\,$\pi,\eta,\rho,\omega,K,K^*$\cite{14}& & 5.4 & & 2.1\\
\hline
\end{tabular}
\end{center}
\caption{Calculated properties of the nonmesonic decay of ${}^5_\Lambda$He.}
\end{table}
Here again what is important is not so much the agreement of disagreement
with experiment but rather the discrepancies {\it between} the various
calculations
which need to be clarified before any significant confrontation between
theory and experiment is possible.

Before leaving this section, it is important to raise an
additional issue which needs to be resolved before reliable
theoretical calculations are possible---that of the $\Delta I={1\over 2}$
rule.\cite{22}   Certainly in any venue in which it has been
tested---nonleptonic kaon
decay---$K\rightarrow 2\pi,3\pi$, hyperon decay---$B\rightarrow B'\pi$,
$\Delta I={1\over 2}$ components of the decay amplitude are found to
be a factor of twenty or so larger than their $\Delta I={3\over 2}$
counterparts.  Thus it has been natural in theoretical analysis of nonmesonic
hypernuclear decay to make this same assumption.  (Indeed without it the
already large number of unknown parameters in the weak vertices expands
by a factor of two.)  However, recently Schumacher has raised a serious
question about the correctness of this assumption, which if verified will
have serious implications about the direction of future theoretical
analyses.\cite{23}  The point is that by use of very light hypernuclear systems
one can isolate the isospin structure of the weak transition.  Specifically,
using a simple delta function interaction model of the hypernuclear
weak decay process, as first written down by Block and Dalitz in 1963\cite{24},
one determines
\begin{eqnarray}
{}^4_\Lambda \mbox{He}: \gamma_4=\Gamma_{NM}(n/p)={2R_{n0}\over 3R_{p1}+R_{p0}}
\nonumber\\
{}^5_\Lambda \mbox{He}: \gamma_5=\Gamma_{NM}(n/p)={3R_{n1}+R_{n0}\over
3R_{p1}+R_{p0}}
\nonumber\\
\gamma=\Gamma_{NM}({}^4_\Lambda \mbox{He})/\Gamma_{NM}({}^4_\Lambda \mbox{H})=
{3R_{p1}+R_{p0}+2R_{n0}\over 3R_{n1}+R_{n0}+2R_{p0}}
\end{eqnarray}
where here $R_{Nj}$ indicates the rate for N-stimulated hypernuclear decay
from an initial configuration having spin j.  One can then isolate the ratio
$R_{n0}/R_{p0}$ by taking the algebraic combination
\begin{equation}
{R_{n0}\over R_{p0}}={\gamma\gamma_4\over 1+\gamma_4-\gamma\gamma_5}
\end{equation}
and from the experimental values\cite{25}
\begin{equation}
\gamma_4=0.27\pm 0.14,\qquad \gamma_5=0.93\pm 0.55,\qquad
\gamma=0.73^{+0.71}_{-0.22}
\end{equation}
we determine
\begin{equation}
{R_{n0}\over R_{p0}}={0.20^{+0.22}_{-0.12}\over 0.59^{+0.80}_{-0.47}}
\end{equation}
in possible conflict with the $\Delta I={1\over 2}$ rule
prediction---$R_{n0}/R_{p0}=2$.\footnote{Note that final state nn or np
configurations which arise from initial ${}^1S_0$ states are of necessity
I=1.}  If confirmed by further theoretical and
experimental analysis this would obviously have important ramifications for
hypernuclear predictions.  However, recent work at KEK has indicated that the
correct value for $\gamma$ should be nearer to unity than to the value 0.73
used above in which case the ratio is considerably increased and there may
be no longer any indication of $\Delta I={1\over 2}$ rule violation.\cite{26}

\section{-- Conclusions}

We have given a brief overview of the field of weak hypernuclear physics.
Because of limited experimental data and of the difficulty of doing
reliable theory, the present situation is quite unsatisfactory.  Although
there is very rough qualitative agreement between theoretical expectations and
experimental measurements, it is not clear whether discrepancies which do
exist are due to experimental uncertainties, to theoretical insufficiencies,
or both.  On the theoretical side, what is needed are reliable calculations
on finite hypernuclei
(preferably by more than one group) which clearly indicate what signals
should be sought in the data.  The issue associated with the validity
of the $\Delta I={1\over 2}$ rule must be clarified.  In addition there have
been recent speculations about the importance of two-nucleon stimulated
decay\cite{27} (which could account for as much as 15\% of the decay amplitude
according so some estimates) and of the importance of direct quark ({\it i.e.}
non-meson-exchange) mechanisms,\cite{28}
which deserve further study in order to eliminate the vexing double
counting problems which arise when both direct quark and meson exchange
components are included.   On the experimental side, we require
an extensive and reliable data base developed in a variety of nuclei in order
to confirm or refute the predicted patterns.  Clearly the strong program
of hypernuclear physics at DA$\Phi$NE will provide a major step in this
direction.

\section*{Acknowledgements}

The above research was supported in part by the National Science Foundation and
by the Department of Energy.  It is also a pleasure to acknowledge the warm
hospitality shown by our hosts at DA$\Phi$NE.


\begin{thebibliography}{99}
\bibitem{1} Part. Data Group, Phys. Rev. {\bf D50}, 1221 (1994).
\bibitem{2} W. Cheston and H. Primakoff, Phys. Rev. {\bf 92}, 1537 (1953).
\bibitem{3} A comprehensive review of such processes is given by J. Cohen,
Prog. Part. Nucl. Phys. {\bf 25}, 139 (1990); see also B.R. Holstein, {\it Weak
Interactions in Nuclei}, Princeton Univ. Press, Princeton (1989), Ch. 5.
\bibitem{4} K.J. Nield et. al., Phys. Rev. {\bf C13}, 1263 (1976).
\bibitem{5} R. Grace et al., Phys. Rev. Lett. {\bf 55}, 1055 (1985);
P.D. Barnes, Nucl. Phys. {\bf A450}, 43c (1986).
\bibitem{6} A comprehensive review of such processes is given by J. Cohen,
Prog. Part. Nucl. Phys. {\bf 25}, 139 (1990); see also B.R. Holstein, {\it Weak
Interactions in Nuclei}, Princeton Univ. Press, Princeton (1989), Ch. 5.
\bibitem{7} J.P. Bocquet et al., Phys. Lett. {\bf B182}, 146 (1986) and
{\bf B192}, 312 (1987); S. Polikanov et al., Nucl. Phys. {\bf A478}, 805c
(1988).
\bibitem{8} B. Desplanques, J.F. Donoghue and B.R. Holstein, Ann. Phys. (NY)
{\bf 124}, 449 (1980).
\bibitem{9} See, e.g., M.M. Nagels et al., Phys. Rev. {\bf D12}, 744 (1975) and
{\bf D15}, 2547(1977).
\bibitem{10} G. Barton, Nuovo Cimento {\bf 19}, 512 (1961).
\bibitem{11} J.B. Adams, Phys. Lett. {\bf 22}, 463 (1966); Phys. Rev. {\bf
156},
1611 (1967).
\bibitem{12} B.H.J. McKellar and B.F. Gibson, Phys. Rev. {\bf C30}, 322 (1984).
\bibitem{13} E. Oset and L.L. Salcedo, Nucl. Phys. {\bf A443}, 704 (1985) and
{\bf A450}, 371c (1986).
\bibitem{14} J.F. Dubach, Nucl. Phys. {\bf A450}, 71c (1986); also, in Proc.
Int. Symp. on Weak and Electromagnetic Int. in Nuclei (WEIN), ed. H.V. Klapdor,
Springer-Verlag (1986); J.F. Dubach et al., UMass preprint (1995).
\bibitem{15} G. Nardulli, Phys. Rev. {\bf C38}, 832 (1988).
\bibitem{16} E.G. Adelberger and W. Haxton, Ann. Rev. Nucl. Part. Sci.
{\bf 35}, 501 (1985).
\bibitem{17} E. Golowich, J.F. Donoghue and B.R. Holstein, Phys. Rept.
{\bf 131}, 319 (1986).
\bibitem{18} B.F. Gibson, Nuovo Cim. {\bf 102A}, 367 (1989).
\bibitem{19} T. Kishimoto, {\it Properties and Interactions of Hyperons}, ed.
B.F. Gibson, P.D. Barnes and K. Nakai, World Scientific, New York (1994).
\bibitem{20} A. Ramos et al., Phys. Lett. {\bf B264}, 233 (1991);
Nucl. Phys. {\bf A544}, 703 (1992) and {\bf A547},
103c (1992).
\bibitem{21} K. Takeuchi, H. Takaki and H. Bando, Prog. Theo. Phys.
{\bf 73}, 841 (1985); H. Bando, Prog. Theo. Phys. Suppl. {\bf 81}, 181 (1985).
\bibitem{22} See, {\it e.g.}, E. Golowich, J.F. Donoghue and B.R. Holstein,
{\it Dynamics of the Standard Model}, Cambridge Univ. Press, New York (1992),
Ch. VIII.
\bibitem{23} R.A. Schumacher, Nucl. Phys. {\bf A547}, 143c (1992).
\bibitem{24} M.M. Block and R.H. Dalitz, Phys. Rev. Lett. {\bf 11}, 96 (1963).
\bibitem{25} V.J. Zeps and G.B. Franklin, to be published in Proc. INS
Symposium \#23, Tokyo (1995).
\bibitem{26} H. Outa, private communication.
\bibitem{27} M. Ericson, Nucl. Phys. {\bf A547}, 127c (1992); M. Shmatikov,
Phys. Lett. {\bf B337}, 48 (1994).
\bibitem{28} See, e.g., C.Y. Cheung, D.P. Heddle and L.S. Kisslinger, Phys.
Rev.
{\bf C27}, 335 (1983); D.P. Heddle and L.S. Kisslinger, Phys. Rev. {\bf C23},
608
(1986); K. Maltman and M. Shmatikov, Phys. Lett. {\bf B331},1 (1994);
T. Inoue, S. Takeuchi and M. Ota, Tokyo Inst. Tech. preprint
TIT/HEP-279/NP (1995).
\end{thebibliography}
\end{document}